\documentclass[smallcondensed]{svjour3}     
\smartqed  
\usepackage{graphicx}
 \usepackage{mathptmx}      
\usepackage{latexsym}

 \journalname{Climatic Change}
\begin{document}


\title{How persistent is civilization growth?}


\author
{Timothy J. Garrett }

\institute{Department of Atmospheric Sciences\\
University of Utah\\
Salt Lake City, USA\\
\email{tim.garrett@utah.edu.}
}


\date{}


\maketitle
\begin{abstract}
In a recent study \cite{GarrettCO2_2009}, I described theoretical
arguments and empirical evidence showing how civilization evolution
might be considered from a purely physical basis. One implication
is that civilization exhibits the property of persistence in its growth.
Here, this argument is elaborated further, and specific near-term
forecasts are provided for key economic variables and anthropogenic
CO$_{2}$ emission rates at global scales. Absent some external shock,
civilization wealth, energy consumption and carbon dioxide emissions
will continue to grow exponentially at an average rate of about 2.3\%
per year. 
\end{abstract}

\section{Introduction}

Through combustion, carbon dioxide (CO$_{2}$) is emitted as a by-product
of the primary energy consumption that is used to run the economy
\cite{Raupach2007}. Anthropogenic CO$_{2}$ emissions accumulate
in the atmosphere \cite{KeelingWhorf2005}, and are a primary control
of changes in global mean climate \cite{Lacis2010}. 

Studies of the response of the atmosphere to changing greenhouse concentrations
are informed by a mixture of observations and a basic understanding
of underlying processes. The evidence is that about 40\% of emitted
carbon remains in the atmosphere \cite{Joos1996,LeQuere2003,Knorr2009}.
Numerical and theoretical models, combined with paleoclimate data,
point to an equilibrium surface temperature response to a doubling
of CO$_{2}$ concentrations that lies somewhere between 2 $^{\circ}$C
and 4.5 $^{\circ}$C \cite{IPCC_WG12007}. 

Meanwhile, economic scientists consider the evolution of civilization
and its emissions to be driven by decisions made by individuals, organizations
and governments \cite{IPCC_WG32007}. The judgement is that human
perceptions and behavior control the rate at which civilization consumes
fossil energy. Policy guides sources of primary energy, rates of human
reproduction, individual wealth and lifestyles, and how efficiently
energy is consumed to produce economic output \cite{Raupach2007}.
Global CO$_{2}$ emission trajectories are determined by these choices.

Unfortunately, there is an exceptionally broad range of CO$_{2}$
emission trajectories that is considered to be humanly plausible,
and this greatly amplifies the uncertainty in the physics \cite{Stott2002}.
Arguably, this is a real problem, especially if climate change becomes
a negative feedback on economic growth \cite{IPCC_WG22007}. If human
adaptation to climate change is to be anything more than purely responsive,
constrained forecasts of global CO$_{2}$ emission trajectories will
certainly be needed. 

In a recent paper in this journal \cite{GarrettCO2_2009}, I suggested
that predictability might be greatly improved if, like climate systems,
human systems were also approached from a physical viewpoint. To this
end, I proposed a thermodynamically-based framework for the evolution
of civilization wealth and its rate of energy consumption at globally
integrated scales. At the core of the prognostic model is a hypothesis
that the instantaneous rate of primary energy consumption by civilization
$a$ is linked through a constant $\lambda$ to its inflation-adjusted
economic value (or civilization wealth) $C$, where wealth is the
time-integral of global economic production (or GDP) $P$, adjusted
for inflation at market exchange rates (MER), and aggregated over
the entirety of civilization history \cite{Maddison2003,UNstats}\begin{equation}
a=\lambda C=\lambda\int_{0}^{t}P\left(t'\right)dt'\label{eq:alambdaC}\end{equation}
Taking $a$ to be in units of Watts, and $P$ in units of 1990 MER
US dollars per second, then wealth $C$ has units of 1990 MER US dollars,
and the constant $\lambda$ has units of Watts per 1990 MER US dollar. 

While this formulation is highly unorthodox from traditional economic
standpoints (see a discussion in Appendix B of Garrett (2011)),
it is nonetheless transparent, and therefore easy to test. What was
found was that for the period 1970 to 2005 for which global statistics
for $a$ were available \cite{AER2009}, the mean value of $\lambda$
amounts to 9.7 milliwatts per 1990 US dollar, with an uncertainty
in the mean at the 95\% confidence level of just 0.3 milliwatts per 1990 US
dollar. 

It appears then that $\lambda$ is indeed constant with time. This
is the empirical support behind the initial hypothesis expressed in
Eq. \ref{eq:alambdaC}, that real global economic value is an expression
of the global capacity to consume primary energy resources. More recent
data extending to 2008 has not changed the value of the derived result
(see Table \ref{tab:Measured-values} and the supplementary material).
If anything, the inter-annual variability in calculated values of
$\lambda$ is diminishing with time%
\begin{table}[htp]
\caption{\emph{\small \label{tab:Measured-values}Measured values for the global
rate of energy consumption $a$ (TW), global real wealth $C$ (trillion
1990 MER USD)}{\small ,}\emph{\small{} the ratio $\lambda=a/C$ (mW
per 1990 MER USD), global real GDP $P$ (trillion 1990 MER USD per
year) and the real growth rate $\eta=P/C$ (\% per year). }}

\begin{tabular}{cccccccccc}
\hline 
 & \textsf{\small 1970} & \textsf{\small 1975} & \textsf{\small 1980} & \textsf{\small 1985} & \textsf{\small 1990} & \textsf{\small 1995} & \textsf{\small 2000} & \textsf{\small 2005} & \textsf{\small 2008}\tabularnewline
\hline
{\footnotesize $a$ } & {\footnotesize 7.2} & {\footnotesize 8.4} & {\footnotesize 9.6} & {\footnotesize 10.3} & {\footnotesize 11.7} & {\footnotesize 12.2} & {\footnotesize 13.2} & {\footnotesize 15.3} & {\footnotesize 16.4}\tabularnewline
{\footnotesize $C=\int_{0}^{t}P\left(t'\right)dt'$ } & {\footnotesize 821} & {\footnotesize 884} & {\footnotesize 960} & {\footnotesize 1048} & {\footnotesize 1151} & {\footnotesize 1266} & {\footnotesize 1398} & {\footnotesize 1536} & {\footnotesize 1656}\tabularnewline
{\footnotesize $\lambda=a/C$ } & {\footnotesize 8.8} & {\footnotesize 9.4} & {\footnotesize 10.0} & {\footnotesize 9.8} & {\footnotesize 10.2} & {\footnotesize 9.6} & {\footnotesize 9.4} & {\footnotesize 9.9} & {\footnotesize 9.9}\tabularnewline
{\footnotesize $P$} & {\footnotesize 11.5} & {\footnotesize 13.9} & {\footnotesize 16.8} & {\footnotesize 19.2} & {\footnotesize 22.3} & {\footnotesize 24.8} & {\footnotesize 29.3} & {\footnotesize 33.6} & {\footnotesize 37.1}\tabularnewline
{\footnotesize $\eta=P/C$ (\% per year)} & {\footnotesize 1.40} & {\footnotesize 1.57} & {\footnotesize 1.75} & {\footnotesize 1.83} & {\footnotesize 1.93} & {\footnotesize 1.96} & {\footnotesize 2.09} & {\footnotesize 2.17} & {\footnotesize 2.24}\tabularnewline
\hline
\end{tabular}

\end{table}
.

\section{Precision versus predictability in economic quantities}

Here, the implications of $\lambda$ being constant for long-range predictability are discussed in greater detail.
The main implication is that global civilization has inertia. Eq. 1 shows that the current
rate of energy consumption $a$ is intrinsically determined by the
entirety of past economic productivity $P$, which, when adjusted
for inflation, yields our current global wealth $C$. Because the past is unchangeable, 
 civilization will carry its current wealth into the future,
and also its associated rate of energy consumption $a=\lambda C$.
Unless there is very rapid decay from some severe external shock,
near-term reductions in energy consumption and wealth are physically
implausible. They would require civilization to somehow {}``forget''
its past accumulation of wealth $C$. 

In general, the variance of any externally forced system demonstrates
the property of {}``reddening'', meaning that it is the most slowly
evolving components of the system that exhibit the most power. The
analogy that could be drawn is to a growing child, or in fact any
other organism \cite{Montieth2000,Brown2004}. Whether the child
is growing or shrinking, energy must still be consumed to sustain
all the internal circulations that have developed through prior growth
of body mass. Accident or a disease could rapidly change rates of
energy consumption through sickness and death. But otherwise, the
child will tend to follow a slowly evolving growth trajectory.

In the same manner, civilization as a whole consumes energy in order
to sustain the material flows that enable it to survive. The current
capacity to consume has evolved from the activities of our ancestors,
through their creation of us, as well as their construction of farms,
towns, communication networks and machines. This past production and
consumption continues to enable us to consume. And, since civilization
is currently very large, it is this accumulated past that will most
strongly govern our future energy consumption and emission rates of
carbon dioxide. 

The growth rate of civilization and its energy consumption can be
expressed in a variety of ways, all of which follow from Eq. \ref{eq:alambdaC}:
\begin{equation}
\eta=\frac{d\ln a}{dt}=\frac{d\ln C}{dt}=\frac{P}{C}=\frac{P}{\int_{0}^{t}P\left(t'\right)dt'}\label{eq:eta}\end{equation}
There is currently no fundamental theory for describing what controls
the evolution of $\eta$ in civilization. However, the data indicate that the growth rate $\eta\left(t\right)$
evolves slowly itself. In 2008 it reached a historical high of 2.24
\% per year (Table \ref{tab:Measured-values}), up from 1.93\% per
year in 1990. It is probably a safe bet to assume that that similar growth rates will
persist in the near-term.

The point here is that persistence is an effective tool for forecasting, but most especially when applied to "reddened" variables like global wealth $C=\int^t_0 Pdt'$, that are highly integrated over time and space. When predicting the evolution of any system, there is always a trade-off. Integral quantities are easier to forecast, but at the sacrifice of temporal and spatial resolution. Specifically, integration biases variability in $P$ towards its more slowly varying components. If $\int_{0}^{t}Pdt'$ is much larger than $P\Delta t$, then even wild short-term fluctuations in $P$
can have only limited impact on the total integral over time. Indeed,
as shown in Table \ref{tab:Measured-values}, this is the scenario
we currently experience. Annual GDP ($P\Delta t$) is only about 2\%
of total global wealth ($\int_{0}^{t}Pdt'$). So even in an artificial scenario where GDP were to suddenly halve for the next five years, it would not have a large impact on global wealth and rates of energy consumption.

The value of examining globally and temporally integrated quantities was a point that was largely missed in two critiques that appeared with Garrett (2011). Cullenward et al. (2011) and Scher and Koomey (2011) argued that there cannot be a constant relationship between energy consumption rates $a$ and wealth $C$ because the relationship between $a$ and  $P$ is highly dynamic, both temporally and between sectors/nations.

This misrepresents the arguments
in Garrett (2011) because the discussion of Eq. \ref{eq:alambdaC}
in Garrett (2011) was explicitly referenced, not to nations
or economic sectors, but to civilization as a whole. More importantly, Eq. \ref{eq:alambdaC} does
not apply to $P/a$, but rather to the integral quantity $C/a = \int_{0}^{t}Pdt'/a$. Certainly, there has
been past discussion among economists that there exists a strong correlation
between rates of energy consumption and economic production at the
national level (e.g., \cite{Costanza1980}). However, $P$ and $\int_{0}^{t}Pdt'$
are not at all the same thing, and they have no obvious relationship to
one another. They might be statistically correlated, but only if $P$
is growing exponentially. 

In fact, if $\lambda = a/C$ is constant, then it is gains in {}``energy productivity'' $f=P/a$ that drive global economic growth. Essentially, $f$ is a measure
of the capacity of civilization to turn current energy consumption
into its own growth, where growth enables civilization to consume
more energy in the future. Combining Eqs. \ref{eq:alambdaC} and \ref{eq:eta},
one finds that the growth of wealth and energy consumption follow
\begin{equation}
\eta=\lambda f=\frac{1}{a}\frac{da}{dt}=\frac{d\ln C}{dt}\label{eq:etaf}\end{equation}
Thus, contrary to what is normally assumed \cite{Raupach2007}, higher
energy productivity corresponds to accelerated growth in energy consumption
rates (see also \cite{Jevons1865,Saunders2000,Owen2010}). Similarly,
the growth in real global GDP follows\begin{equation}
\frac{d\ln P}{dt}=\eta+\frac{d\ln\eta}{dt}\equiv\lambda f+\frac{d\ln f}{dt}\label{eq:dlnPdt}\end{equation}

For example, the mean value of $f=P/a$ between 1970 and 2008 was
61 micro-dollars per joule, where dollars are expressed in inflation-adjusted
1990 MER units. Thus, the mean value of $\lambda f$ for this period
is 5.9$\times$10$^{-8}$ \% per second or 1.87 percent per year.
A linear least-squares fit to the observed trend in $f$ is $d\ln f/dt=$
1.00 \% yr$^{-1}$. Thus, the thermodynamically based model provides
a mean calculated growth rate for world real GDP between 1970 and
2008 of $d\ln P/dt=$ 1.87 $+$ 1.00 $=$ 2.97 \% per year. The actual
observed value based on a least-squares fit to the data is 2.90\%
per year. The difference between observations and theory is just two
percent. 

Such accuracy in the global GDP growth calculation is really just a consequence
of there being a constant factor $\lambda$ relating global wealth
$\int_{0}^{t}P\left(t\right)dt'$ to energy consumption rates $a$
(Table \ref{tab:Measured-values}); the rest is just basic math. But,
perhaps since $P$ increased by more than a factor of three between
1970 and 2008, this analysis might lend further reassurance that the
model is empirically validated, and that it can provide simplified
forecasts of global GDP growth without having to resort to explicit
representations of nations, sectors, people or their lifestyles%
\footnote{In fact, it is the absence of people in the model that is the justification
for evaluating fiscal quantities in units of MER rather than purchasing
power parity (PPP) currency, as has been advised \cite{Cullenward2010}. %
}.

\section{Persistence in growth}

To reiterate, available statistics show that wealth, when it is integrated over the entire global economy and integrated over the entire history of economic production, has been related to the current rate of global primary energy consumption through a factor that  has
been effectively constant over nearly four decades of civilization growth. Its implications for the future are that civilization has inertia, and therefore its current rate of consumption
growth is unlikely to cease in a hurry.

 From Eq. \ref{eq:eta}, $\eta$ is the rate of growth of real wealth
$C$ and energy consumption rates $a$. The value of $\eta$ is intrinsically
tied to the energy productivity $f=P/a$ through Eq. \ref{eq:etaf}.
Thus, the rate of growth in $\eta$ itself, or $d\ln\eta/dt$, can
be termed as the {}``real innovation rate'' since it corresponds
to greater energy productivity. In an innovative world, the deterministic
solution for the growth of energy consumption rates $a$, wealth $C$
and CO$_{2}$ emission rates $E$ (assuming they stay tied to energy
consumption rates) is of form \cite{GarrettCO2_2009}

\begin{equation}
X\left(t\right)=X_{0}e^{\eta\tau_{\eta}\left(e^{t/\tau_{\eta}}-1\right)}\label{eq:X-1}\end{equation}
where $X$ refers to any of $a$, $C$, or $E$. $X_{0}$ is some
initial condition and $\tau_{\eta}$ is the time constant for growth
in $\eta$ \begin{equation}
\tau_{\eta}=\frac{1}{d\ln\eta/dt}\label{eq:taua-1}\end{equation}
Note that the solution for $X\left(t\right)$ (Eq. \ref{eq:X-1})
condenses to the simple exponential growth form of $X=X_{0}\exp\eta t$
in the limit that innovation slows to the point that $\tau_{\eta}\rightarrow\infty$.
If there is innovation, however, then $\tau_{\eta}$ is positive and
finite, and growth is super-exponential (i.e. the exponent of an exponent). 

Another way of expressing innovation and its relationship to growth
is to think of doubling-times. The doubling times $\delta$ for wealth
$C$ and growth rates $\eta$ are given by \begin{equation}
\delta_{X}=\frac{\ln2}{\eta}\label{eq:doubX}\end{equation}
\begin{equation}
\delta_{\eta}=\frac{\ln2}{d\ln\eta/dt}\label{eq:doubeta}\end{equation}

Effectively $\delta_{X}$ represents the time required for civilization
to double its wealth at current rates, and $\delta_{\eta}$ is the
time required for the growth rate to double (or $\delta_{X}$ to halve).
Thus, from Eq. \ref{eq:X-1}, a deterministic solution for growth
in wealth, energy consumption or CO$_{2}$ emissions follows\begin{equation}
X\left(t\right)=X_{0}2^{\frac{\delta\eta}{\delta_{X}}\left(2^{t/\delta_{\eta}}-1\right)}\label{eq:X2}\end{equation}

\begin{figure}
\includegraphics[width=11cm]{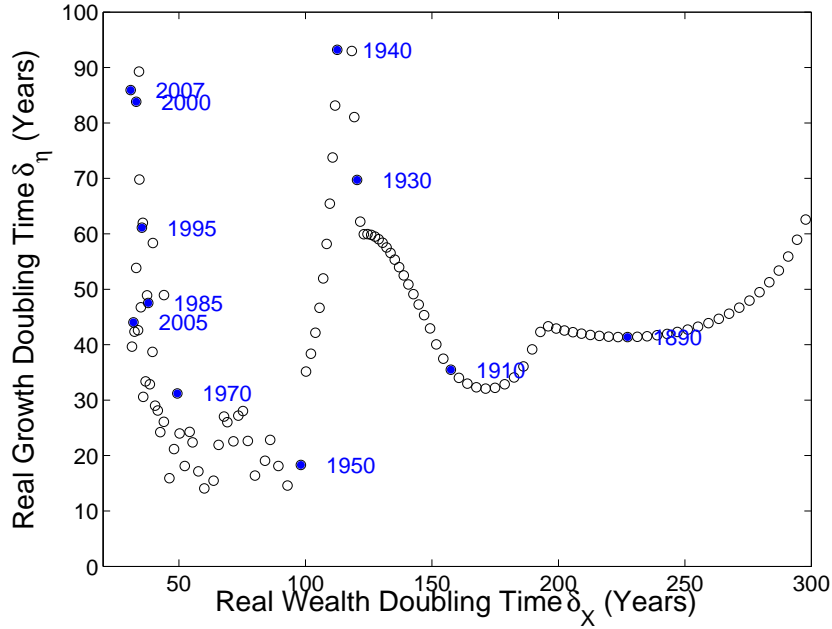}

\caption{\label{fig:doubling}Values of the doubling time for the growth rate
of wealth $\delta_{\eta}$ (Eq. \ref{eq:doubeta}) versus the doubling
time for wealth itself $\delta_{X}$ (Eq. \ref{eq:doubX}). Select
years are shown for reference. }

\end{figure}

Historical statistics for $\delta_{X}$ and $\delta_{\eta}$ are shown
in Fig. \ref{fig:doubling}. For the purpose of the calculations,
the definition of exponential growth rates is $\eta=P/C$ (Eq. \ref{eq:eta}).
The data show that, over the past century or so, there has been a long term tendency for
wealth to double over ever shorter intervals. As a whole, the world is getting richer faster. 

More interesting than the growth rate, however, is the innovation time, which itself
shows marked inflection points. There were two periods of particularly
rapid innovation. The first was in the late
1800s and early 1900s, when the growth rate doubled in 40 years. Following
1950, the growth rate doubled over a remarkably short timespan of
just 20 years. %

For contrast, both the 1930s and the past decade have been characterized
by more stagnant innovation, or a relaxation from super-exponential
to simple exponential growth. Currently, wealth is doubling more quickly
than ever before; it now takes only about 30 years for global wealth
to double. Still, for the first time since the Great Depression, the
doubling time has nearly ceased to shorten.

It is interesting to speculate as to what might be driving the variability
in $\delta_{\eta}$. Perhaps it is access to important new oil reservoirs
in the late 1800s and around 1950 that led to bursts in innovation,
and an absence of large discoveries is slowing real innovation today.
If so, it hints at a more fundamental thermodynamic theory for civilization
evolution that incorporates the accessibility and depletion of geological
reservoirs. 

In the meantime, one can take advantage of recent stagnation in innovation
to make simplified near-term forecasts for the growth of the human
system and its CO$_{2}$ emissions (Table \ref{tab:forecasts}). Assuming
persistence, on average, global energy consumption rates and wealth
will continue to grow at a rate of about 2.3\% per year, or a doubling
time of 30 years. Assuming that global decarbonization continues to
be extremely slow \cite{Raupach2007}, the same rate will apply to
global emissions of CO$_{2}$, $E$. From Eq. \ref{eq:dlnPdt}, global
real GDP will grow at a rate that is only marginally faster than the growth rate for wealth, since
the innovation rate $d\ln\eta/dt$ is approaching zero. %
\begin{table}
\caption{{\small \label{tab:forecasts}Forecasted growth rates for energy consumption
rates $a$, wealth $C$, global production $P$ and carbon dioxide
emission rates $E$. The value $\epsilon$ is a quantity much smaller
than the associated rate. Financial quantities are with respect to
inflation-adjusted MER units. }}

\noindent \centering{}\begin{tabular}{ccccc}
\hline 
 & {\small $a$ } & {\small $C$} & {\small $P$} & {\small $E$}\tabularnewline
\hline
{\small Growth rate (\% yr$^{-1}$)} & {\small 2.3} & {\small 2.3} & {\small 2.3$+\epsilon$} & {\small 2.3}\tabularnewline
{\small Doubling time (yr)} & {\small 30} & {\small 30} & {\small 30$-\epsilon$} & {\small 30 }\tabularnewline
\hline
\end{tabular}
\end{table}

As a strong word of caution, persistence is never something to carry
too far. As a guess, the rates provide in Table \ref{tab:forecasts}
apply only for timescales significantly less than the wealth doubling
time of 30 years. Essentially the future is unknowable, and unforeseen
catastrophes or boons cannot be excluded. Equally, exponential growth
cannot continue unabated because, sooner or later, civilization must
face up to resource depletion or environmental degradation.

\section{Conclusions}

Sometimes one sees the naive argument that climate scientists are
presumptuous to make long-range forecasts of climate when short-term
weather forecasts are so often wrong. What makes climate forecasts
possible is top-down energetic constraints. It is not necessary to explicitly model weather
in order to make long-term forecasts of globally-averaged
surface temperatures. With certain assumptions about relative humidity
and clouds, the key ingredients for a simple 1D radiative-convective
equilibrium climate model are the rate of solar energetic input, and
the concentration of greenhouse gases \cite{ManabeWetherald1967}.
Long-range predictions of regional climate variability require greater
sophistication. But even here, top-down constraints dictate the plausible
range of climatolological parameter space \cite{IPCC_WG12007}. 

Scher and Koomey (2011) have argued that {}``Economic systems are not the same
as physical systems, and we shouldn\textquoteright{}t model them as
if they are''. Nonetheless, civilization is part of the physical
universe. As with climate and weather, its evolution should also be constrained by global scale energetic flows. The evidence presented here and in
Garrett (2011) suggests that it is indeed possible to make
long-term forecasts of global energy consumption rates, without having
to explicitly model the internal, short-term details of people and
their lifestyles. Long-range forecasts of energy consumption by specific
countries or economic sectors will be more difficult \cite{Scher2010}
and require additional sophistication. But, from the standpoint of
determining emission rates of a long-lived gas such as CO$_{2}$,
the internal details are largely irrelevant. So long as there is atmospheric mixing and
international trade, it is only global scale energy consumption and CO$_{2}$ emissions that matter. 

The main point made in Garrett (2011) was that the global
economy can be placed on a physical footing, through a constant coefficient
that links economic wealth (not wealth production) to the global consumption
rate of primary energy resources. The relevant physics is still too
primitive to provide a fully deterministic solution into the future.
Still, as argued here, one can apply the principle of persistence based
on recent trends, provided one is looking at quantities that are highly integrated over space and time. Just as one might make the purely statistical argument
that recent trends in globally-averaged surface temperatures will continue into the near future,
here it is suggested that near-term growth in economic wealth and
energy consumption rates will also persist. The qualification is that the
growth will not be super-exponential, as it has been in past decades,
but more purely exponential. The forecasted growth rate is about
2.3 $\%$ per year. 

\newpage

\bibliographystyle{spmpsci} 
\bibliography{References}

\begin{thebibliography}{10}
\providecommand{\url}[1]{{#1}}
\providecommand{\urlprefix}{URL }
\expandafter\ifx\csname urlstyle\endcsname\relax
  \providecommand{\doi}[1]{DOI~\discretionary{}{}{}#1}\else
  \providecommand{\doi}{DOI~\discretionary{}{}{}\begingroup
  \urlstyle{rm}\Url}\fi

\bibitem{IPCC_WG32007}
{Climate Change 2007 - Mitigation of Climate Change}.
\newblock {Cambridge University Press} (2007)

\bibitem{AER2009}
{Annual Energy Review 2009}.
\newblock Tech. Rep. DOE/EIA-0384(2009), {Department of Energy, Energy
  Information Administration} (2009).
\newblock \urlprefix\url{www.eia.doe.gov/aer/inter.html}

\bibitem{UNstats}
{United Nations Statistical Databases} (2010).
\newblock \urlprefix\url{unstats.un.org/unsd/snaama}

\bibitem{Brown2004}
Brown, J.H., Gillooly, J.F., Allen, A.P., Savage, V.M., West, G.B.: Toward a
  metabolic theory of ecology.
\newblock Ecology \textbf{85}, 1771--1789 (2004)

\bibitem{Costanza1980}
Costanza, R.: Embodied energy and economic valuation.
\newblock Science \textbf{210}, 1219--1224 (1980)

\bibitem{Cullenward2010}
Cullenward, D., Schipper, L., Sudarshan, A., Howarth, R.B.: Psychohistory
  revisited: fundamental issues in forecasting climate futures.
\newblock Clim. Change \textbf{3}, 457--472 (2011)

\bibitem{GarrettCO2_2009}
Garrett, T.J.: Are there basic physical constraints on future anthropogenic
  emissions of carbon dioxide?
\newblock Clim. Change \textbf{3}, 437--455 (2011).
\newblock \doi{10.1007/s10584-009-9717-9}

\bibitem{IPCC_WG22007}
IPCC: {Climate Change 2007 - Impacts, Adaption and Vulnerability}.
\newblock Cambridge University Press (2007)

\bibitem{IPCC_WG12007}
IPCC: {Climate Change 2007 - The Physical Basis}.
\newblock Cambridge University Press (2007)

\bibitem{Jevons1865}
Jevons, W.S.: {The Coal Question}.
\newblock Macmillan and Co. (1865)

\bibitem{Joos1996}
{Joos}, F., {Bruno}, M., {Fink}, R., {Siegenthaler}, U., {Stocker}, T.F., {Le
  Qu{\'e}r{\'e}}, C., {Sarmiento}, J.L.: {An efficient and accurate
  representation of complex oceanic and biospheric models of anthropogenic
  carbon uptake}.
\newblock Tellus B \textbf{48}, 397--417 (1996)

\bibitem{KeelingWhorf2005}
Keeling, C.D., Whorf, T.P.: Trends: A Compendium of Data on Global Change,
  chap. {Atmospheric CO$_{2}$ records from sites in the SIO air sampling
  network}.
\newblock Carbon Dioxide Information Analysis Center, Oak Ridge National
  Laboratory, U.S. Department of Energy, Oak Ridge, Tenn., U.S.A. (2005)

\bibitem{Knorr2009}
{Knorr}, W.: {Is the airborne fraction of anthropogenic CO$_{2}$ emissions
  increasing?}
\newblock Geophys. Res. Lett. \textbf{36}, L21,710 (2009).
\newblock \doi{10.1029/2009GL040613}

\bibitem{Lacis2010}
Lacis, A.A., Schmidt, G.A., Rind, D., Ruedy, R.A.: Atmospheric co$_2$:
  Principal control knob governing earth's temperature.
\newblock Science \textbf{330}, 356--359 (2010).
\newblock \doi{10.1126/science.1190653}

\bibitem{LeQuere2003}
{Le Qu{\'e}r{\'e}}, C., {Aumont}, O., {Bopp}, L., {Bousquet}, P., {Ciais}, P.,
  {Francey}, R., {Heimann}, M., {Keeling}, C.D., {Keeling}, R.F., {Kheshgi},
  H., {Peylin}, P., {Piper}, S.C., {Prentice}, I.C., {Rayner}, P.J.: {Two
  decades of ocean CO$_{2}$ sink and variability}.
\newblock Tellus B \textbf{55}, 649--656 (2003).
\newblock \doi{10.1034/j.1600-0889.2003.00043.x}

\bibitem{Maddison2003}
Maddison, A.: {The World Economy: Historical Statistics}.
\newblock OECD (2003)

\bibitem{ManabeWetherald1967}
Manabe, S., Wetherald, R.T.: Thermal equilibrium of the atmosphere with a given
  distribution of relative humidity.
\newblock J. Atmos. Sci. \textbf{24}, 241--259 (1967)

\bibitem{Montieth2000}
Montieth, J.L.: Fundamental equations for growth in uniform stands of
  vegetation.
\newblock Agricult. Forest. Meteorol. \textbf{5-11} (2000)

\bibitem{Owen2010}
Owen, D.: The efficiency dilemma.
\newblock The New Yorker pp. 78--85 (2010)

\bibitem{Raupach2007}
Raupach, M.R., Marland, G., Ciais, P., Le~Qu\'{e}r\'{e}, C., Canadell, J.G.,
  Klepper, G., Field, C.: {Global and regional drivers of accelerating CO$_{2}$
  emissions}.
\newblock Proc. Nat. Acad. Sci.  (2007).
\newblock \doi{10.1073/pnas.0700609104}

\bibitem{Saunders2000}
Saunders, H.D.: {A view from the macro side: rebound, backfire, and
  Khazzoom-Brookes}.
\newblock Energy Policy \textbf{28}, 439--449 (2000)

\bibitem{Scher2010}
Scher, I., Koomey, J.: {Is accurate forecasting of economic systems possible?
  An editorial comment}.
\newblock Clim. Change \textbf{3}, 473--479 (2010).
\newblock \doi{0.1007/s10584-010-9945-z}

\bibitem{Stott2002}
Stott, P.A., Kettleborough, J.A.: Origins and estimates of uncertainty in
  predictions of twenty-first century temperature rise.
\newblock Nature \textbf{416}, 723--725 (2002)

\end{thebibliography}

\end{document}